# Hierarchical macro-nanoporous metals for leakage-free high-thermal conductivity shape-stabilized phase change materials

Yaroslav Grosu,[1,*] Yanqi Zhao,[2] Alberto Giacomello,[3] Simone Meloni,[3,4] Jean-Luc Dauvergne,[1] Artem Nikulin,[1] Elena Palomo,[1,5] Yulong Ding,[2] Abdessamad Faik[1]

[1] *Centre for Cooperative Research on Alternative Energies (CIC energiGUNE), Basque Research and Technology Alliance (BRTA), Alava Technology Park, Albert Einstein 48, 01510 Vitoria-Gasteiz, Spain ygrosu@cicenergigune.com*

[2] *BCES Birmingham Centre of Energy Storage, University of Birmingham, United Kingdom*

[3] *Dipartimento di Ingegneria Meccanica e Aerospaziale, Sapienza Università di Roma, via Eudossiana 18, 00184 Rome, Italy*

[4] *Dipartimento di Scienze Chimiche e Farmaceutiche (DipSCF), Università degli Studi di Ferrara (Unife), Via Luigi Borsari 46, I-44121, Ferrara, Italy*

[5] *Ikerbasque, Basque Foundation for Science, 48013 Bilbao, Spain*

## Abstract

**Impregnation of Phase Change Materials (PCMs) into a porous medium is a promising way to stabilize their shape and improve thermal conductivity, which are essential for thermal energy storage and thermal management of small-size applications, such as electronic devices or batteries. However, in these composites a general understanding of how leakage is related to the characteristics of the porous material is still lacking. As a result, the energy density and the antileakage capability are often antagonistically coupled. In this work we overcome the current limitations, showing that a high energy density can be reached together with superior anti-leakage performance by using hierarchical macro-nanoporous metals for PCMs impregnation. By analyzing capillary phenomena and synthesizing a new type of material, it was demonstrated that a hierarchical trimodal macro-nanoporous metal (copper) provides superior antileakage capability (due to strong capillary forces in nanopores), high energy density (90 vol% of PCM load due to macropores) and improves the charging-discharging kinetics, due to a three-fold enhancement of thermal conductivity. It was further demonstrated by**

**CFD simulations that such a composite can be used for thermal management of a battery pack and, unlike pure PCM, it is capable of maintaining the maximum temperature below the safety limit. The present results pave the way for the application of hierarchical macro-nanoporous metals for high-energy density, leakage-free, and shape-stabilized PCMs with enhanced thermal conductivity. These innovative composites can significantly facilitate the thermal management of compact energy systems such as electronic devices or high-power batteries by improving their efficiency, durability, and sustainability.**

1. Introduction

Future energy systems will be comprised of a combination of different sources, which include fossil and nuclear fuels used in large plants with renewable energies produced in micro-generators distributed over the territory (solar panels, micro-wind turbines) [1]. Due to a strongly discontinuous energy output of this ecosystem, energy storage devices are a key ingredient. In this scenario a one-size-fits-all approach is not suitable since large energy storage technologies (such as pumped-storage hydropower) are expected to work in parallel with smaller storage devices, which will be increasingly needed to support local energy generation. Other elements of future energy ecosystems will be devices for energy scavenging, to recover energy that otherwise will be dispersed in the environment, and thermal energy management to keep primary energy storage devices to work at optimal temperatures. As will be discussed in detail, there are devices combining all these functionalities in one apparatus.

Phase change materials (PCMs) currently represent an attractive solution for thermal energy storage (TES) and thermal management applications where a high energy density in a narrow temperature range is required [1,2]. Most commonly, heat is stored as latent melting heat – the enthalpy that is absorbed by a system during the solid–liquid phase transition [2]. A great variety of materials were explored for this purpose, covering a wide range of temperatures starting from near room temperature to temperatures as high as 900 °C [3].

Organic PCMs, such as saturated fatty [4] and carboxylic [5] acids, sugar alcohols [6,7], amides [8] and alkanes [9], are promising PCMs in the low-temperature range. Salt hydrates PCMs were also considered for vehicle thermal management [10], domestic heating [11] and building energy storage [12]. However, corrosion, thermal stability [13], and supercooling[a] [14,15] prevent a safe and efficient energy recovery. For high- and medium-temperature thermal energy storage inorganic salts [16,17] and alloys are typically considered [18]. Despite many advantages of PCMs, their possible leakage is a serious engineering limitation, which can lead to PCM loss and damage of the equipment. Moreover, PCMs often suffer from low thermal conductivity increasing their charge–discharge time.

In order to overcome the leakage and low thermal conductivity problems, a class of materials called shape-stabilized PCMs was introduced. Their use is beneficial for applications such as hybrid heating systems [19], buildings [20,21,22], electronic components [23], Li-ion batteries thermal management [24,25], photovoltaic modules [26], thermoregulating textiles [27,28], solar-driven cookers [29], water heaters [30] and refrigerators [31]. Different composites were used for this purpose such as binary fatty acid/diatomite [32], aluminum foam/ paraffin [33], $NaNO_3/Ca(OH)_2$ [34], MgO/carbonate salt [35], polyethylene glycol/silica gel/β-Aluminum nitride [36], polyethylene glycol/expanded graphite [37]. Some commercial products are available on the market, such as microencapsulated PCMs, which are produced by several manufacturers (Pure PCM, Rubitherm GmbH, Andores New Energy Co., Ltd. Shanghai Tempered Entropy New Energy Co., Microtek Laboratories Inc.) for applications such as textile fabrics, heat storage boards, heat pouches for therapeutic use, cooling vests, heat therapy pack, construction, electronics cooling, building and packaging [38]. However, both laboratory and commercial materials have some limitations that will be discussed below.

Thermal management of batteries is one of the potential applications of shape-stabilized PCMs. It is known that both the performance and the lifetime of a battery are highly dependent on temperature. In particular, temperatures above 60 °C are considered non-safe for battery packs. Shape stabilized PCMs are attractive for the thermal management of batteries due to their low operating cost and facile integration. However, leakage as well as thermal expansion problems must be mitigated [38]. Lv et al. [39] addressed these issues by proposing 4-component composite, consisting of paraffin, expanded graphite (EG), low density polyethylene (PE) and a small amount of nanosilica. It was demonstrated that nanosilica stabilized the liquid phase paraffin and limited the expansion. Tomizawa et al. [40] developed a thermal management system for mobile devices by using microencapsulated PCM and PE composite in the form of sheets, while melamine resin was used to encapsulate the paraffin. In that work it was demonstrated that PCM can be used for efficient thermal management of electronic devices and that using thick PCM sheet along with highly thermal conductive copper sheets could effectively limit the temperature rise without influencing the transistors saturation times, a transistor property determining the time it takes them to switch off.

---

[a] In some PCM, on the timescale of the application solidification takes place at a temperature that is lower from the *thermodynamic* transition temperature. This is related to the presence of free energy barriers between the solid and liquid state. The need to supercool the liquid to extract the latent heat stored reduces the utility of the material and, if too severe, prevents heat recovery [15].

Shape-stabilized PCMs typically imply blending of PCM with solid particles [37] or impregnation of solid–liquid PCM into a porous material/ foam with high thermal conductivity [41]. For example, Rehman and Ali studied thermal performance of copper and Iron-Nickel foams impregnated with paraffin for thermal management of electronics [42]. The authors reported considerable thermal conductivity enhancement and base temperature reduction [42]. Recently Chen et al. proposed carbon-based in situ reduction of metal ions for development of highly conductive matrix for PCMs impregnation [43]. The authors applied this method to copper microspheres introduced into a carbon-based adsorbent and then impregnated the resulting matrix with PEG, demonstrating a stable composite with more than two-fold thermal conductivity increase compared to pure PEG [43]. The use of porous materials and foams was recently reviewed by Rehman at al. [44]. Carbon black nanoparticles were explored to develop shape-stabilized PCMs with enhanced thermal conductivity as well as photo-thermal conversion using auric acid [45], paraffin wax [46] and phenol-water system doped with different nanoparticles [47]. A 3D graphene was used to enhance the thermal conductivity of paraffin and to improve its shape stabilization [48]. More than twofold enhancement of thermal conductivity of sodium thiosulfate pentahydrate was demonstrated by adding carbon nano-tubes [49]. Cellulosic scaffolds doped with boron nitride nanosheets were used to develop shape-stabilized yet flexible phase change composites based on polyethylene glycol with 42.8% enhanced thermal conductivity [50]. An interesting approach to enhance heat transport in PCMs by exploiting the thermocapillary effect was recently proposed [51,52]. With this solution, there are no mass penalization nor issues related to thermal expansion. It was shown that, in n-octadecane PCM, the thermal Marangoni effect increases heat transfer by approximately two times [52].

Nanoporous materials were identified as promising candidates to mitigate the leakage of PCMs due to strong capillary forces provided by nanopores [53], which was demonstrated using metal–organic framework [54], 3D porous carbon [55] and porous scaffolds of boron nitride [56]. In this sense, a new class of porous materials – nanoporous metals – looks particularly attractive due to their high thermal conductivity. Such materials can be obtained as plasmonic metamaterials [57] or by dealloying [58], i.e. a selective corrosion of one or several metals forming the alloy [59]. The range of reported nanoporous metals is very wide and includes porous gold obtained through electrochemical [60,61] and chemical [62,63] dealloying of Al-based alloys, palladium obtained through electrochemical [64,65] or chemical [66] dealloying of Al-based alloys and electrochemical dealloying of Pd30Ni50P2 [67] alloys. Similarly, nanoporous platinum [64], silver [68], nickel [69–71] and copper [72] were reported. The range was extended to some composites like AuPt [53,73], AuPd, AuPtPd [53], Si-Ag [74], PdAu [75], YNiCo [76].

A limitation of available nanoporous metals is that they have a low volumetric porosity as compared to macroporous materials and foams, which reduces the overall energy density and capacity of the PCM – porous metal composite. On the other hand, the small size of the pores would induce large capillary forces, thus effectively hindering leakage. This reasoning highlights an obvious conflict between the energy capacity and the antileakage capability of PCM – porous metal composites. In this work we demonstrate that this conflict can be solved and that it is actually possible to engineer leakage-free, high-energy density PCM – porous metal composites by exploiting hierarchical macro-nanoporous structuring of the metal substrate. In particular, we demonstrate both theoretically and experimentally that a hierarchical pore morphology allows to take advantage both of macroporosity, which allows for high energy density and capacity (high volume of impregnated PCM), and of nanoporosity, which allows to prevent leakage due to the strong capillary forces. Simultaneously, the high thermal conductivity of the porous metal results in an enhancement of the thermal conductivity as compared to PCM alone.

To prove the concept, we developed a new trimodal macro-nanoporous copper substrate obtained by dealloying Mg-Cu-Zn alloy of close-to-eutectic compositions. In particular, 63 wt% Mg – 22 wt% Cu – 15 wt% Zn alloy was used (the eutectic composition is 60 wt% Mg – 25 wt% Cu – 15 wt% Zn). The use of close-to-eutectic composition allows a considerable increase of the porosity, reaching 90 vol%. It should be noted that the proposed approach of dealloying close-to-eutectic alloys is not specific to copper and can be applied to a great variety of known porous metals. The substrate was impregnated with paraffin and the high energy density, enhanced thermal conductivity and antileakage properties of the shape-stabilized PCM material are demonstrated. A case study simulation was also preformed, demonstrating the suitability of the proposed system for battery pack thermal management to improve their efficiency and durability.

Present results open a new route for developing high-energy density, leakage-free and shape-stabilized PCMs with enhanced thermal conductivity by exploiting the properties of hierarchical macro-nanoporous metals.

The article is structured as follows. In Section 2, materials and methods used for this work are described. In Section 3, the theory linking leakage with the effect of roughness is provided. The main results are discussed in Section 4, which includes hierarchical macro-nanoporous copper preparation, characterization, and impregnation (Section 4.1), anti-leakage performance of Cu+paraffin composite and its stability under thermal cycling (Section 4.2), thermal conductivity measurements (Section 4.3) and CFD modeling of a battery pack, where the proposed composite was used for the thermal management (Section 4.4). The supporting results are provided in Supplementary materials. The summary of the article is presented in the Conclusions section.

## 2. Materials and methods

*2.1 Materials.*

Cu, Mg and Zn (99.9%) were purchased from Alfa Aesar. Acetic acid and paraffin (CAS 8002–74-2, melting temperature 53–58 °C) were purchased from Sigma-Aldrich. The chemicals were used as received.

Alloys preparation. The use of a ternary alloy was expected to provide a higher porosity as compared to the binary system [72]. With this regard, 63 wt% Mg – 22 wt% Cu – 15 wt% Zn alloys were prepared as follows. Bulk 3–5 mm spherical pieces of Mg, Zn and Cu were polished and placed in an alumina crucible. The polishing was performed with an ATM Saphir stationary polishing machine using abrasive 2500 SiC discs under flowing water. The rotation speed was set to 100 rpm and a polishing time of 10 mins was used for each sample. Alumina crucible with the metals was inserted into an alumina tube, which was closed from both sides using alumina plugs with an inlet and outlet for argon ventilation. The alumina tube with the crucible was placed into a tubular furnace. A constant flow of argon was blown through the tube during the synthesis. The temperature of the furnace tube was raised at a rate of 10 °C/min and held at a temperature of 750 °C for 2 h. The samples were then cooled at a rate of 10 °C/min. After this temperature treatment, the sample was turned upside down and heated a second time with the same conditions to reach homogeneity. The weight of the sample before and after the synthesis was compared to exclude the possibility of partial Mg evaporation at elevated temperatures.

As was demonstrated in our previous work [72], in order to obtain macropores one needs to shift from the eutectic composition, providing the excess of metal expected to be dealloyed. Following this principle, we considered the close-to-eutectic composition 63 wt% Mg – 22 wt% Cu – 15 wt% Zn alloy, where excess of Mg was used to create macropores at the expense of Cu (the eutectic composition is 60 wt% Mg – 25 wt% Cu – 15 wt% Zn). Other close-to-eutectic compositions were explored. However, there is a trade-off between porosity and mechanical stability of the final material [72]. The chosen composition resulted to be the best compromise between these two characteristics.

*2.2. Dealloying process and paraffin impregnation*

The synthesized Mg-Cu-Zn alloy samples were shaped by cutting and polishing in the form of discs or rectangular pieces with characteristic size of 10 mm and thickness of 0.8–1.0 mm. These pieces were subjected to dealloying by exposure to acetic acid under free corrosion conditions: the samples were immersed into the acid and kept at room temperature and atmospheric pressure. The maximum immersion time was 16 h. After dealloying, the samples were washed in water and ethanol, followed by thermal treatment at 200 °C for 4 h under argon atmosphere to remove the acid from the pores and to cure the porous structure.

Completely dealloyed samples (confirmed by X-ray diffraction (XRD) and energy dispersive x-ray spectroscopy (EDX)) were subjected to vacuum impregnation with paraffin as follows. The porous sample was placed in a glass tube along with a bulk piece of paraffin. Next, the tube was outgassed under vacuum of $5 \cdot 10^{-3}$ mbar for 2 h. Then, the temperature of the system was raised to 80 °C, while maintaining the vacuum. This temperature increase led to the paraffin melting and its adsorption into the outgassed pores of copper. Next, while maintaining the temperature at 80 °C, the atmospheric pressure was re-established in the glass tube. Finally, the impregnated sample was removed from the tube.

*2.3. Scanning electron microscopy (SEM)*

The samples were imaged by a scanning electron microscope, namely a Quanta 200 FEG, which was operated in high vacuum mode at 5–30 kV using a backscattered electron detector (BSED) and Everhart-Thornley Detector (ETD). In addition, EDX analysis was carried out for elemental mapping. As-synthesized alloys were polished successively with SiC discs, followed by polishing using diamond nanoparticles suspension prior to the SEM imaging. The nanoparticles suspension was used for polishing as it provides superior smoothness of the samples as compared to SiC abrasive discs. This smoothness is essential for high-quality SEM imaging. Diamond nanoparticles were chosen as carbon is not present in the alloy composition, so any possible contamination with nanoparticles can be easily identified by EDX analysis. To avoid such contamination, after the polishing the samples were cleaned in an ultrasonic bath using acetone, ethanol and water for 30 mins each. This ensured that no contamination with nanoparticles occurred (as verified by EDX) and that SEM results were not affected by polishing. Dealloyed samples were imaged without further preparation. Additionally, the dealloyed samples were carefully broken into several pieces for examination of the cross-section.

*2.4. X-ray diffraction (XRD)*

Bruker D8 Advance diffractometer equipped with a LYNXEYE detector using CuKα1 radiation (λ = 1.5418 A) and θ–2θ geometry was used for XRD analysis. Data were collected at room temperature between 25° and 80° with a step size of 0.02° and a counting time of 8 s per step. The EVA software was used to determine the phase composition of the material.

*2.5. The flash method*

This method proposed by Parker et al. [77] was used to determine thermal diffusivity. The method consists in subjecting one face of a sample of finite thickness to a short-duration energy pulse. The thermal response versus time of the opposite side was then recorded and processed to estimate the thermal diffusivity of the sample. In this work, the samples were subjected to a radiative energy supplied by a 150 W halogen lamp. The corresponding thermal responses were recorded using a FLIR A6752SC infrared camera (30 Hz, during c.a. 1 min). The samples have dimensions of approximately 10x10x1 mm. For this experiment samples were introduced into an opaque insulating material and the measurement was conducted along the 10 mm dimension. Three tests with different solicitations (radiative energy supplied during 1.3–2.6 s) were performed for each sample.

The experimental thermal responses were processed by minimizing their difference with a 1D model using a Levenberg-Marquardt fitting method [78]. The model, based on the quadrupole formalism [79], takes into account the convective heat losses on the front and rear faces. A pre-treatment of the data allowed the finite duration of the excitation as well as the semitransparency of the samples to be taken into account by introducing a correction based on the center of gravity of the energy pulse and an offset, respectively. The algorithm of minimization was initialized considering a thermal diffusivity pre-estimated by the method proposed by Degiovanni [80] and assuming the absence of convective heat losses.

*2.6. Nitrogen adsorption*

Textural properties were characterized in an automated gas adsorption analyzer (Micromeritics ASAP 2460). Nitrogen sorption isotherms of the samples were measured after outgassing at 200 °C in a vacuum for 5 h. The multipoint surface area was evaluated with the Brunauer-Emmett-Teller (BET) method over the range $P/P_0 = 0.075–0.35$ and pore size distribution was obtained using Barrett- Joyner-Halenda (BJH) model applied to the desorption isotherm branch. The total pore volume was determined from the volume adsorbed at $P/P_0 = 0.98$.

*2.7. Thermogravimetric analysis*

Thermogravimetry was performed using the NETZSCH STA 449 F3 Jupiter thermal analyzer at a constant argon flow of 60 mL/min in the temperature range from 25 °C to 600 °C with the heating rate of 10 °C/ min.

*2.8. Differential Scanning Calorimetry (DSC)*

Specific heat capacity ($C_p$), enthalpy, and melting temperature of paraffin and Cu + paraffin composite were measured by the Discovery DSC 2500 from TA instruments, which allows a direct $C_p$ measurement in the 10 °C/min ramp mode. The instrument was calibrated using sapphire and indium. Aluminum hermetic pans were used as reference as well as sample crucibles. They were hermetically closed under argon atmosphere and controlled humidity inside the glove box. Before and after each measurement, the specific heat capacity of sapphire was measured in the same temperature range to validate the uncertainty of the measurement. Three measurements were performed for each sample and 3 different samples of paraffin as well as of Cu + paraffin composites were measured. The uncertainty provided in Table 1 for these properties is calculated as the standard deviation from the average value.

*2.9. Thermal cycling*

The prepared Cu + composites were tested for thermal stability under thermal cycling conditions. The samples were introduced into the Carbolite CWF 11/5 laboratory chamber furnace, which was programmed for 10 heating cooling cycles at 2 °C/min. Such heating rate was chosen as this is the maximum achievable cooling rate for this equipment. The tests were conducted under air atmosphere. Temperature in close proximity of the samples was controlled using an external thermocouple connected to a digital temperature controller. This was done to ensure that the samples experience the required temperatures, since in general the temperature distribution in the furnace can be non-homogeneous. After 10 heating–cooling cycles the samples were removed to measure their mass using analytical balances. After that the protocol was repeated 4 times to reach 50 cycles in total.

## 3. Theory

In this section, we discuss the effect of nanoporosity on paraffin impregnation in and leakage from the porous matrix. We try to provide a general discussion, which applies to all PCMs and porous matrices and that could be useful for designing shape-stabilized, leakage-free PCMs.

The first problem, which does not appear to be fully explained in the literature, concerns the driving force of leakage. In the following, we analyze three hypotheses and we discuss how the presence of nanoporosity can influence (limit) leakage.

A first hypothesis was that gravity, acting on the impregnating fluid, increases the pressure at the bottom pores and is able to cause a fluid leak outside of the pores. However, a back of the envelope calculation comparing the order of magnitude of the pressure due to gravity, $\rho g h \approx 10 \; Pa$ - where $\rho$ is the PCM density, $g$ is the acceleration of gravity, and $h$ is the sample height of the order of 1 mm - and the capillary pressure $2\gamma_{lv}/R \approx 100$ kPa - where $\gamma_{lv}$ is the liquid-vapor surface tension and $R$ is the radius of the largest pores, of the order of 1 μm - immediately reveals that capillarity is always many orders of magnitude larger than gravity, and thus the latter does not play a crucial role in leakage in this kind of systems.

The second possible driving force of leakage is the thermal expansion of the PCM during the phase transition, which may drive some liquid paraffin out of the pores. The overall thermal expansion of paraffin can be evaluated as a 10 % increase over the initial solid volume. However, this kind of leakage cannot explain the approximately 50 % loss of PCM observed in several shape-stabilized PCM systems [81], and could be overcome simply by supplying a smaller initial volume of PCM or by providing a means of absorbing the excess paraffin during the first cycles.

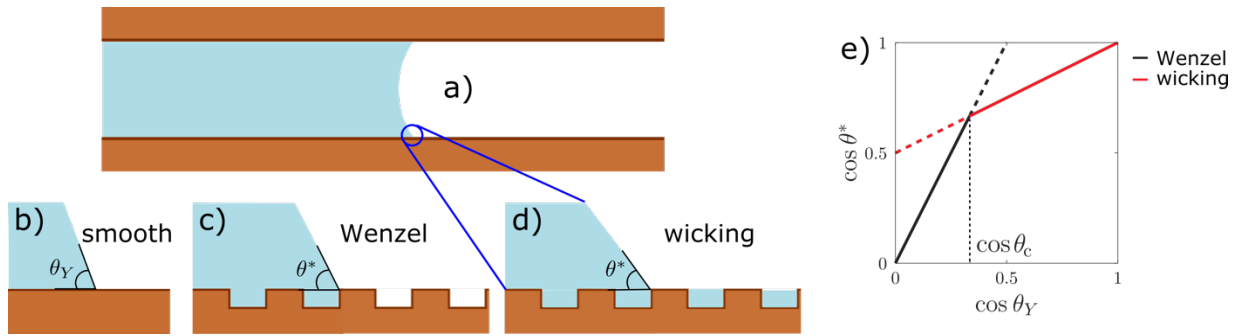

**Figure 1.** Liquid penetrating in a hydrophilic pore (a) with smooth (b) or rough (c-d) walls. e) Apparent contact angle $\theta^*$ in the Wenzel and wicking states as a function of the Young contact angle $\theta_Y$; $\theta_c$ is the critical angle, at which the apparent angles related to the states are the same.

The third origin of leakage is to be sought in the nucleation and expansion of gas bubbles within the pores occurring during the temperature cycles. Typically, during the impregnation process only the porous system is outgassed, while PCM can still trap a significant volume of air. In addition, air can penetrate in the pores during thermal cycling. This air can coalesce and expand during the operation, forming air bubbles that can cause significant leakage of PCM. Consequently, a significant reduction of the performance and durability of the system can be the end result. For a gas bubble to form and grow and, thus push the liquid paraffin out of the microchannels, the bubble must overcome an energy barrier – the nucleation barrier [82]. This occurs by *thermal fluctuations* of the bubble, i.e. fluctuations of its energy that normally occur in a system at constant temperature. Thermal energy fluctuations are typically of the order of the thermal energy available to the system, i.e. $k_BT$, where $k_B$ is the Boltzmann constant and $T$ is the temperature of the system. Thus, the formation and growth of bubbles occur only when the barrier is of the order of $k_BT$. Nanopores increase liquid adhesion to the macropores and, as a consequence, increase the nucleation barrier thus preventing nucleation.

A convenient way to measure liquid adhesion is the effective contact angle $\theta^*$, the angle formed by a droplet of the liquid and the apparent, macroscopic surface of the material, encompassing the effect of nanoporosity within $\theta^*$ (Figure 1): the higher the adhesion, the lower the effective contact angle. In the following we summarize the relation between the intrinsic chemical characteristics of the PCM and of the solid material and the degree of porosity of the nanopores. In particular, based on the Wenzel and wicking models of wetting of the porous matrix (Figure 1) it was possible to provide estimates of $\theta^*$. Based on this value we will determine which is the thermodynamically stable state (Wenzel or wicking) and how porosity enhances liquid retention. It was found that the nucleation barrier for forming a bubble grows linearly with $\cos\theta^*$, which establishes a quantitative relation between the chemical and physical characteristics of the material and the leakage resistance.

A second and final effect induced by nanoporosity is that on substrates with hydrophilic surface chemistry, it magnifies contact angle hysteresis. This is the difference between the advancing ($\theta_a$) and receding ($\theta_r$) contact angles [83], the contact angle of the *front* and *rear* menisci of the displacing liquid respectively. This is important for leakage because the work required to displace a liquid tube outside a pore is proportional to $2R\,\gamma_{lv}\,(\cos\theta_r - \cos\theta_a)\,dx$, where *dx* is the infinitesimal displacement of the liquid tube along the axis of a cylindrical pore of radius *R*. We remark that this effect of nanopores is complementary to their capacity to suppress nucleation of gas bubbles and helps preventing leakage in conditions where the nucleation barrier is small or negligible.

The presence of nanopores within the walls of macropores, which are partially filled by a liquid can lead to two distinct wetting states as for the case of a hydrophilic solid [83, 84] (Figure 1). The first corresponds to the so-called Wenzel state, in which the roughness is wet by the liquid (*only*) below the main liquid body (Figure 1c). In this case, the apparent contact angle for the meniscus of the liquid front in the pore is [83, 84]:

$$cos\theta^* = r\,cos\theta_Y, \qquad (1)$$

where $r$ is the so-called solid roughness, i.e., the ratio of the actual solid area (which includes the internal area of nanopores wetted by PCM) over the projected one, $\theta_Y$ is the ideal contact angle characterizing the surface chemistry of the solid, i.e. the contact angle of the liquid on an atomically smooth surface (cf. Figure 1b). The ideal contact angle is determined by the Young's law $\cos\theta_Y = (\gamma_{sv} - \gamma_{sl})/\gamma_{lv}$. If the solid is hydrophilic the presence of roughness magnifies its hydrophilicity leading to $\theta^*_{Wenzel} < \theta_Y$, which was observed in the present case.

A second state exists, which is relevant when large channels are only partially filled with paraffin or when a gas bubble forms. In this state the liquid film imbibes the nanopores before the meniscus (Figure 1d), leading to a "wicking" state with apparent contact angle [83, 84, 85]

$$cos\theta^* = 1 - \varphi_s(1 - cos\theta_Y), \quad (2)$$

where $\varphi_s$ is the solid fraction remaining dry on top of the roughness *before* the liquid front. Also, in this case the effective hydrophilicity of the surface is enhanced by the presence of nanoporosity, leading to $\theta^*_{wicking} < \theta_Y$.

When a channel is only partially filled, both the Wenzel and wicking states can exist. Which one of the two is the stable state of the liquid can be predicted based on energetic arguments: the system will be found in the Wenzel or wicking state, which has the lowest surface energy. This condition can be translated into an equivalent condition on the contact angle, when the Young contact angle is above the critical value $\theta_c = \arccos[(1 - \varphi_s)/(r - \varphi_s)]$ it is expected that the nanopores are in the Wenzel state and below $\theta_c$ to be in the wicked state (Figure 1e). Since $\theta^* < \theta_Y$ regardless of whether the liquid is in the Wenzel or wicked states, the presence of nanopores always increases adhesion.

Overall, the main driving force of leakage is effectively hindered by the presence of a second tier of hydrophilic roughness at the nanoscale because i) it increases the energetic cost required to nucleate bubbles and, even when such bubbles are formed, ii) the work required to displace the liquid outside of the largest pores is significantly higher as compared to the case of smooth walls.

## 4. Results and discussion
### 4.1. Porous copper preparation, characterization and impregnation

The synthesized Mg-Cu-Zn alloys consist of several crystallographic phases as demonstrated in Figure 2. Combining these results with SEMEDX analysis (Figure 3), it was found that the excess of Mg in the initial composition resulted in the formation of a large flower-like inclusions of Mg phase having ~ 20-50 µm size (Figure 3a, 4a and 4d), as expected from our previous work [72]. Such flower-like inclusions were sometimes surrounded by round 1-2 µm inclusions of Mg (Figure 4c). Other intermetallic phases

(namely, CuMgZn, $Cu_2Mg$, $CuMg_2$, $Mg_{21}Zn_{25}$) formed a complicated pattern (Figure 3), which was homogeneously distributed all over the sample along with flower-like inclusions of Mg (Figure 4a, 4c and 4d).

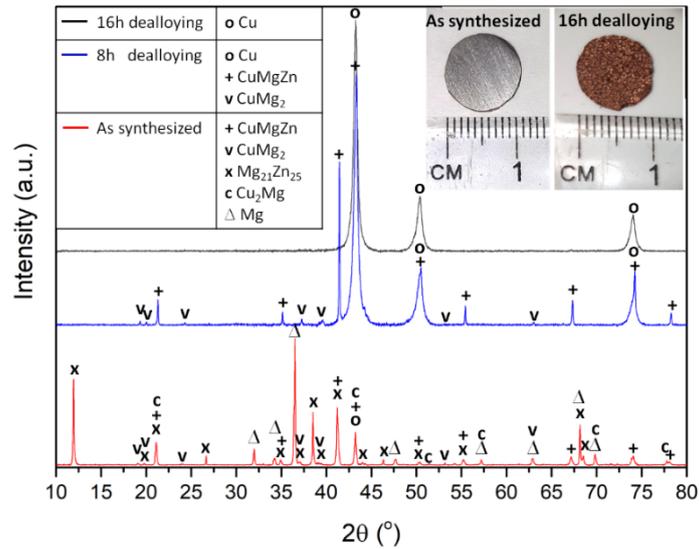

**Figure 2. XRD patterns of as-synthesized Mg-Cu-Zn alloy and after 8 and 16 hours of dealloying in acetic acid. Insets: photos of Mg-Cu-Zn alloy as-synthesized and after 16 hours of dealloying**

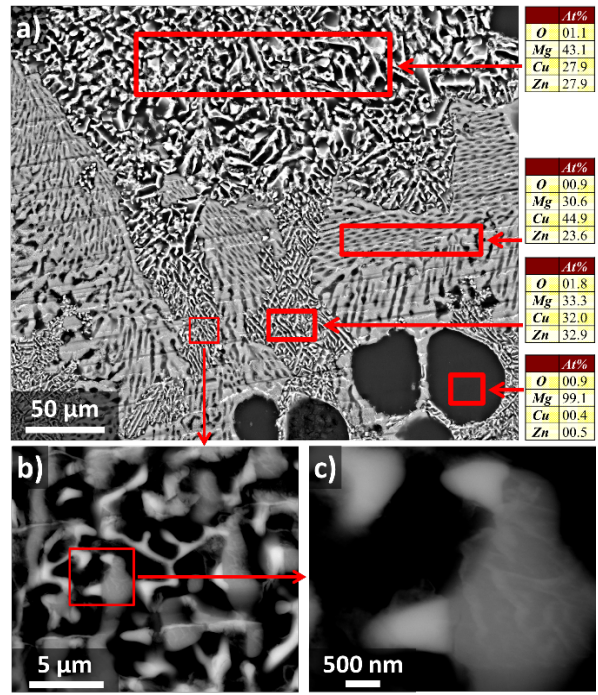

**Figure 3. SEM images and EDX quantitative analysis of as-synthesized Cu-Mg-Zn alloy**

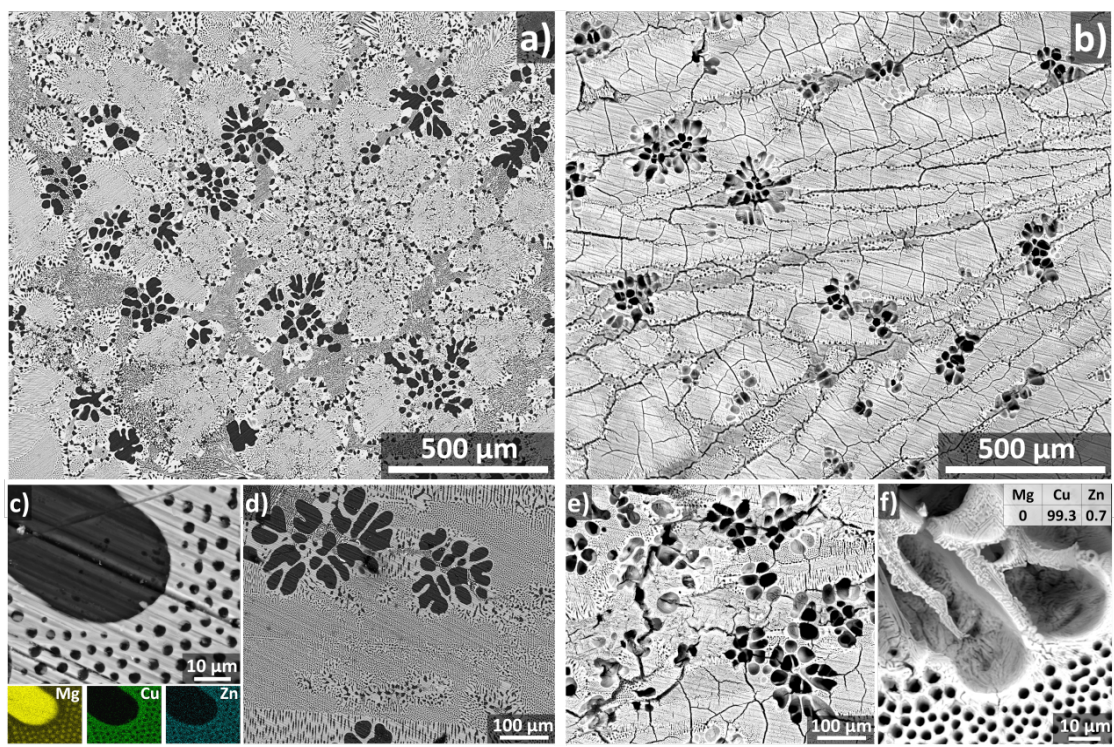

**Figure 4. SEM-EDX analysis of Mg-Cu-Zn alloy after synthesis (panels a, c, d) and after 16 hours dealloying (panels b, e, f)**

As expected from previous articles [64, 65, 72], exposure of alloys consisting of Cu, Mg and/or Zn to acidic media, results in the selective corrosion of Mg and Zn, while the porous Cu structure remains in place. Figure 2 shows that after 8 h of dealloying a pure Cu phase appears in the XRD pattern, while $Cu_2Mg$ and $Mg_{21}Zn_{25}$ phases were no longer present. At the same time, CuMgZn and $CuMg_2$ phases were not completely removed. Dealloying for 8 more hours (overall 16 h) results in the complete removal of Mg- and Zn-containing phases, with just the Cu phase left (Figure 2). After 16-hours of dealloying the samples exhibited the reddish color typical of Cu (insert of Figure 2).

From Figure 4 one can see that all the inclusions of Mg were transformed into pores of corresponding size upon dealloying. High magnification micrographs revealed that three-tier hierarchical porosity was formed upon complete dealloying of Mg- and Zn-containing phases (Figure 5). The largest flower-like inclusions of 20-50 μm as well as the rounded structures surrounding them (1-2 μm) turned into pores of the corresponding size (Figures 5a and 5b). The walls of these pores also contain pores of size ~500-700 nm (Figure 5c). For simplicity we call them medium-size pores. In turn, the medium pores are decorated with ~15-20 nm pores (Figure 5d). This nanoporosity was also characterized by nitrogen gas adsorption experiments, which demonstrated a broad pore size distribution with average pore size of 30 nm (Figure S1, here and below letter "S" signifies that a Figure can be found in Supplementary information file), surface area of (20 ± 3) m$^2$/g and pore volume of (0.12 ± 0.02) cm$^3$/g.

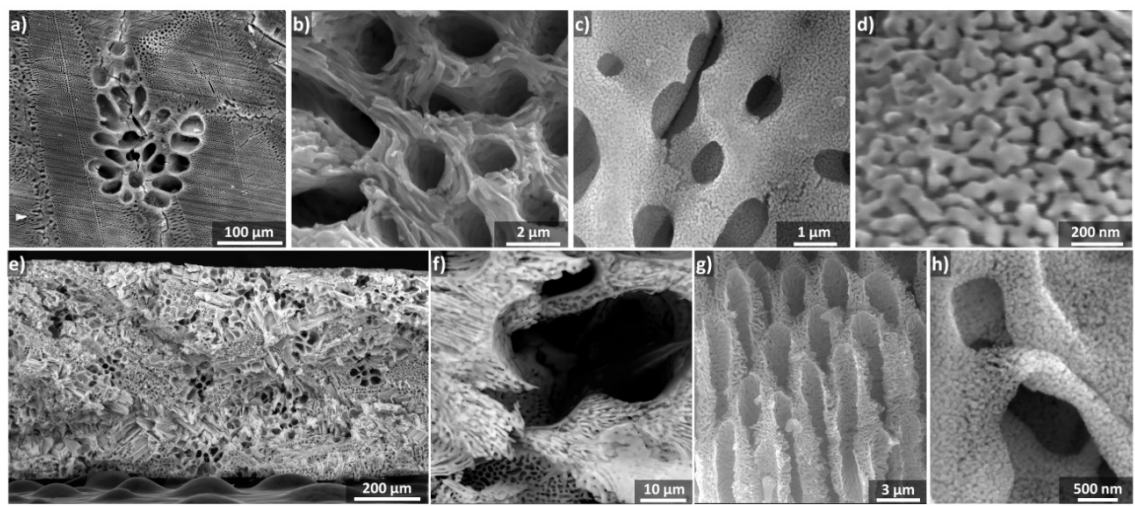

**Figure 5. Surface (panels a, b, c, d) and cross-section (e, f, g, h) views of Mg-Cu-Zn alloy after 16 hours dealloying**

The cross-section examination revealed complete dealloying of the samples of 0.8 – 1.0 mm thickness (Figures 5e-5h). Considering the overall size of the sample, Cu density and its mass, we estimate an overall porosity of ~ 90 vol%. This value was confirmed by comparing the mass of the samples before and after paraffin impregnation (see below and Section 2 of the Supplementary information for more details).

Next, dealloyed samples were vacuum impregnated with paraffin according to the procedure described in Section 2. This resulted in a Cu + paraffin composites with approximately 50 wt% of paraffin, which was identified by comparing the mass of the sample before and after the impregnation, as well as by thermogravimetric analysis (Figure S2) and DSC analysis (Figure 6). More information on the paraffin load estimation is provided in Section 3 of the Supplementary Information. Comparing the mass of paraffin impregnating the porous Cu and the pore volume determined as mentioned above, we conclude that the vacuum impregnation results in high degree of paraffin loading. However, as will be discussed below, some nanopores were not completely impregnated and were filled only after thermal cycling (paraffin migration from the macropores to the nanopores). Therefore, some increase of paraffin load could be achieved by improving the impregnation procedure, e.g. the impregnation time (see discussion below). Any additional sizable increase of paraffin content in the sample can be achieved only by increasing the porosity of the metal sample.

DSC measurements demonstrated that Cu + paraffin composites have a slightly lower melting temperature compared to pure paraffin, namely, 46.67 °C vs 46.96 °C. Such 0.3 °C decrease is close to the statistical uncertainty of the experiment. At the same time, the temperature of the melting peak for the composite increased by 0.6 °C as compared to paraffin (54.38 vs 53.79 °C), which suggests a slight broadening of the peak. Both observations may be related to confinement effects, however, they can be considered negligible for thermal storage applications. The composite stores 51.8 % less enthalpy by mass (Figure 6), which further confirms ~50 wt% of paraffin in the composite. In particular, the enthalpy of 139.4 ± 1.1 J/g was recorded for pure paraffin, while for the composite it was 72.2 ± 0.9 J/g. Considering that the loading of PCM was approximately 50 wt%, a corresponding decrease of the energy capacity (J/g) is expected. The important point for applications, however, is that the energy density (J/m$^3$) decreases only by ~ 10 %. This is because only ~10 vol% of the composite is copper, the remaining ~90 vol% is paraffin. Generally speaking, such a decrease of energy density is unavoidable since a corresponding volume of PCM is substituted by the porous matrix. A loss of 5-15 vol% of PCM is quite typical and considered acceptable [44].

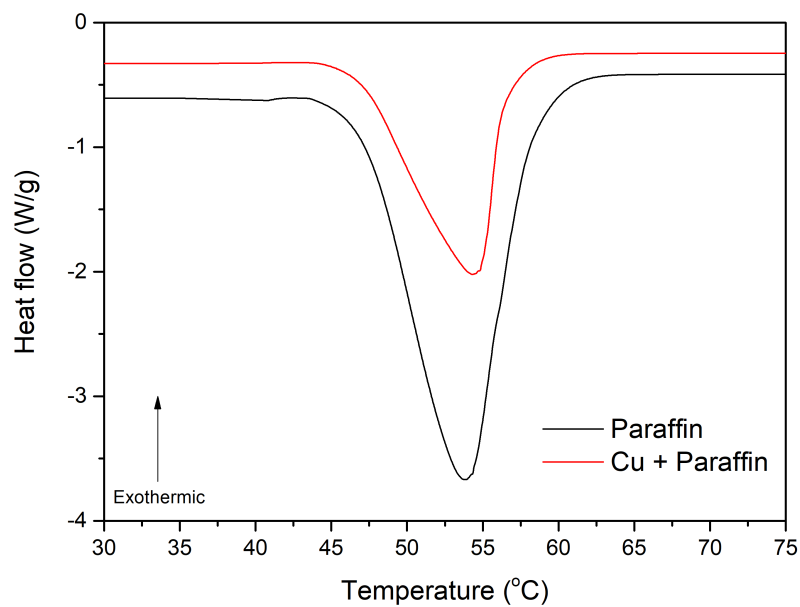

**Figure 6. Differential Scanning Calorimetry curves for Paraffin and Cu + paraffin composite**

## 4.2. Anti-leakage capability and stability of Cu + paraffin composite

As discussed previously, leakage is a major problem in composite PCMs, significantly limiting their application. Thus, the synthesized composites were tested against leakage under thermal cycling as described in Section 2. Figure 7 demonstrates that the Cu + paraffin composite does not leak in its operational temperature range. The mass of the samples remained constant after 50 cycles at 25-75 °C (mass was verified every 10 cycles – Figure 7b) and the visual aspect of the composite did not change (Figure 8a). Together with XRD analysis, which demonstrates the absence of new chemical phases after the cycling (Figure S3), the mass measurement suggests that the composite was stable and did not undergo any degradation or oxidation in the tested temperature range. Additionally, we did not observe any enthalpy change after 50 heating-cooling cycles as seen from DSC analysis (Figure S4). Finally, SEM images demonstrate that there was no cracking of the framework upon thermal cycling (Figure S5). However, noticeable differences can be observed concerning the impregnation of the various levels of porosity by paraffin (Figure 8b and 8c). In particular, just after impregnation, all the pores with the exception of some part of the smallest ones (15-20 nm) were filled – Figure 8b. The smallest pores, which cover the wall of the 0.5 – 2.0 μm pores, were initially filled for only the top ~200 nm of their micrometric thickness. After thermal cycling, some paraffin migrated from the larger pores deeper into the nanoporous walls, completely wetting the nanoporosity between larger pores (Figure 8c). In other words, during the impregnation phase the larger pores were completely filled and the nanometric ones were partially empty because more time is required for the liquid paraffin to percolate into them. As predicted by the theory, with thermal cycles the nanometric pores get completely filled by some of the paraffin originally occupying the large pore, confirming that they served as a reservoir for thermal expansion of the PCM during melting and prevented the paraffin extrusion from the pores. These observations suggest that by increasing the impregnation time (time allowed for paraffin to be sucked by the porous copper under vacuum), higher levels of paraffin impregnation can be reached. Considering that the volume of nanopores constitutes around 10 % of the whole porosity, one may expect of up to around 5 vol% increase of the overall paraffin impregnation, assuming that half of the nanoporosity was filled upon impregnation (estimated from SEM analysis - Figures 8 and S5).

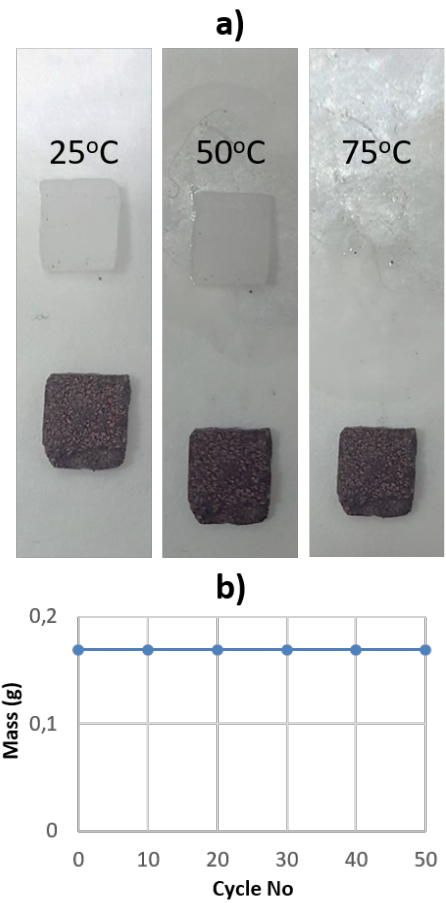

**Figure 7.** a) Photographs of Paraffin (top) and shape stabilized Cu + paraffin composite (bottom) at different temperatures; b) mass variation of the composite during thermal cycling in the 25 – 75 °C temperature range

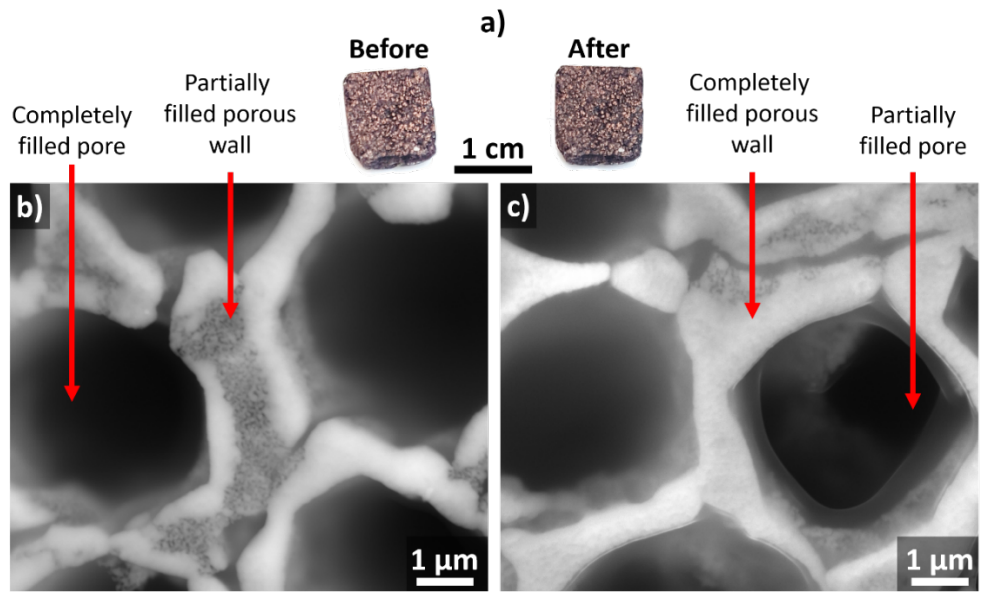

**Figure 8.** a) Photographs and SEM images of Cu + paraffin composite before (b) and after (c) 50 cycles in the 25-75 °C temperature range

Experimental data reported in Figures 5d, 8b and 8c, gas sorption results (Figure S1 in Supplementary information) and the density of the sample allow us to draw some conclusions on the enhanced adherence discussed in Section 3. The Young contact angle of paraffin estimated on a smooth copper surface is $\theta_Y \approx 10°$ (Figure S6). The solid roughness can be obtained from the combination of gas sorption data and the density of the sample. It was estimated to be *r* ~ 15 at the end of the first cycle, when nanopores were imbibed for approximately one third, and *r* ~ 44 after 50 cycles, when nanopores were completely imbibed (see Section 6 and Figure S7 of the Supplementary material for more details on these estimates). From SEM micrographs (Figure 5d) we computed *via* ImageJ [86] the solid fraction of the pore walls due to nanoporosity, i.e. the ratio between the solid (non vacuum) and total lateral area of macropores, which resulted to be $\varphi_s \sim 0.7$. With these positions the critical angle discriminating between the Wenzel and wicking states is $\theta_c \sim 88°$ in the first cycle and ~90° after 50 cycles, meaning that the wicking state is thermodynamically favored and one expects that in the long run the nanoporosity will be wet even when the largest pores are empty. This is consistent with the results shown in Figure 8c, illustrating a partially filled macropore whose nanopores were completely wet by PCM. The presence of nanopores reduces the apparent contact angle to $\theta^* = 8°$ (Eq. 2), making macropores somewhat more wetted by paraffin than what dictated by their chemical nature. This helps stop leakage by increasing the nucleation barrier and preventing the flow of paraffin.

Present results suggest that the hierarchical macro-nano porosity not only provides high energy density (90 vol% of PCM load) due to macropores, but also superior antileakage capability due to the strong capillary forces induced by nanopores, which decorate the walls of the macropores. This is particularly clear in Figure 8, which demonstrates that paraffin preferentially migrates from macropores to nanopores rather than outside of the composite.

### 4.3. Thermal conductivity improvement

In addition to solving the leakage problem, using Cu as the impregnation matrix is also expected to enhance the thermal conductivity of the composite. This is highly beneficial for TES applications, reducing the charging/discharging time and local overheating (see Sec. 4.4). In order to evaluate the thermal conductivity enhancement of the composite as compared to pure paraffin, the flash method was applied according to the procedure described in Section 2. Figure 9a shows the experimental thermograms after the flash impulse for the paraffin and Cu + paraffin samples. One can see that, due to the higher thermal diffusivity, the temperature response of Cu + paraffin sample was noticeably sharper as compared to paraffin. The maximum temperature of the sample was reached ca. 3 times faster, as shown in Figures 9c-9e reporting the thermal scans recorded by the infrared camera at 3 instants of the test. In particular, comparing Figures 9c and 9d it can be concluded that after 12 s the pure paraffin sample did not appreciably change its temperature, while Cu + paraffin has almost reached its maximum. Quantitative results for the thermal diffusivity of Paraffin and Cu + Paraffin samples are summarized in Table S1. Despite the variable radiative exposure times and the noisy signals, the estimated thermal diffusivities show a satisfactory repeatability.

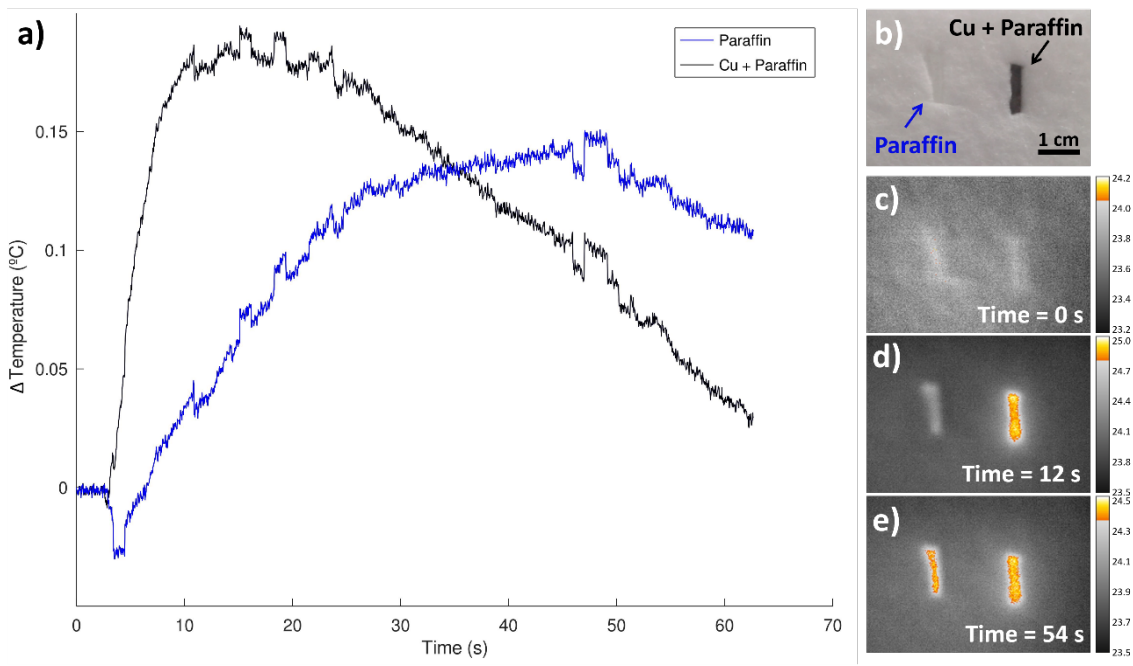

**Figure 9. Flash experiment: a) Experimental thermograms for Paraffin (Blue line) and Cu + paraffin composite (Black line); b) Photographs of the samples placed in a thermal insulator for the experiment; c)-e) Infrared thermal scans at different times during the experiment**

The values obtained for thermal diffusivity $a$ can be used to determine the thermal conductivity via the equation $\lambda = a\rho C_p$, with $\rho$ density of the sample and $C_p$ is its specific heat capacity. However, it should be noted that the required values of $\rho$ and $C_p$ of paraffin deviate quite strongly in the literature [87-90]. In particular, literature results range from $574\ kg/m^3$ [87] to $893\ kg/m^3$ [88] for the density and from $2100\ J/kg \cdot K$ [90] to $4422\ J/kg \cdot K$ [89] for the specific heat capacity. This range of values provides an enhancement of the thermal conductivity for the proposed composite in the range of 3.2 – 3.5. Additionally, the density and $C_p$ values were determined in this work taking into account the broad range of literature values. In particular, $C_p$ of paraffin and Cu + paraffin are shown in Figure S8 and Table 1. Density was calculated using the geometrical dimensions and the weight of rectangular pieces of paraffin (Table 1). Using these values, the thermal conductivity for pure paraffin and the Cu + paraffin composite was calculated (Table 1). It can be seen that the thermal conductivity of the composite is at least three times higher than that of paraffin. However, due to the uncertainty of the obtained values this enhancement actually lies in the range of 3.0 – 6.2.

**Table 1: Thermal diffusivity $(a)$, density $(\rho)$, heat capacity $(C_p)$ and thermal conductivity $(\lambda)$ for Paraffin and Cu + Paraffin samples at 20 °C**

|  | $a$, mm²/s | $\delta a$, mm²/s | $\rho$, kg/m³ | $\delta \rho$, kg/m³ | $C_p$, J/kg·K | $\delta C_p$, J/kg·K | $\lambda$, W/m·K | $\delta \lambda$, W/m·K |
|---|---|---|---|---|---|---|---|---|
| Paraffin | 0.226 | 0.035 | 752 | 37 | 2951 | 103 | 0.50 | 0.12 |
| Cu + paraffin | 0.715 | 0.061 | 1690 | 31 | 1745 | 44 | 2.11 | 0.27 |

Summarizing, our estimate brings us to the conclusion that the Cu + paraffin composite was characterized by a thermal conductivity which is at least 3 times higher than that of pure paraffin. Comparable enhancement of thermal conductivity was obtained by other authors when copper mesh or foams with macroscopic pores on the millimeter scale were used for paraffin impregnation [32, 91].

It is useful to compare metal foams to the hierarchical macro-nanoporous metals proposed in this work in terms of their applicability. While both types of materials can be used for PCM impregnation, they seem to be suitable for a different application niche. Metal foams can be produced as large, thick pieces suitable for large applications on the scale of kilograms. However, a considerable care should be taken in terms of leakage. So typically, PCM-impregnated metal foams are introduced in a hermetic container. On the other hand, hierarchical macro-nanoporous metals proposed in this work provide superior antileakage characteristics due to the strong capillary forces exerted by nanopores, which allows their use in direct contact with the heat source (a battery, for example). However, production of a large samples of these materials may be complicated (at least using the dealloying method) since the dealloying depth is typically quite limited. Therefore, porous metals would be more suitable for small applications where the absence of a protective container is highly beneficial due to space limitations, like compact battery packs and small electronics. The striking advantage of the proposed trimodal hierarchical metals prepared by dealloying alloys of close-to-eutectic compositions as compared to other nanoporous metals explored so far in the literature is their high volumetric porosity due to macropores, and consequently their high energy density.

### 4.4. Performance analysis of the Cu + paraffin composite

In this section we analyze the performance of the Cu + paraffin composite for the thermal regulation of a primary energy storage, a battery pack. As mentioned in the introduction, on top of the direct application for thermal energy storage, shape-stabilized PCM can be used for thermal regulation of primary storages, to keep them working in optimal temperature conditions to maximize their performance, durability and safety. A widely adopted technique to control the temperature of batteries is liquid cooling [92], but the performance of this approach is not completely satisfactory; one can combine liquid cooling and (shape-stabilized) PCM to improve the thermal management of battery packs. In the future it is foreseen that shape-stabilized PCM might completely replace liquid cooling of battery packs [93].

Here, we consider a battery pack composed of 20 lithium ion batteries cooled by a PCM-filled array and a pipe with flowing water (Figure 10). A two-dimensional computational model of this battery thermal management system (BTMS) was established using ANSYS fluent. The model was validated in a previous study [94] and further validated in the present work against numerical and experimental data from Zhao et al. [91]. A full description of the simulation procedure, meshing (Figure S9), validation (Figure S10), input heat profile (Figure S11), and input thermophysical properties (Figure S8 and S12) of the model can be found in the Supplementary information. The model was designed to be compact with battery energy density of 300 Wh/L. The model has dimensions of 90 mm x 137 mm, and gaps between the cells were set to be 3.6 mm in vertical and 7.8 mm in horizontal. The cooling pipes were ribbon-shaped with width of 2 mm. Apart from the heated walls, all the other walls were considered adiabatic, which eliminates additional heat losses. We remark that this is a conservative hypothesis, and one expects that a realistic device is even more efficient. A real-time battery heat generation rate, which was obtained from a previous study [94], was used as input heat source in the model. Different inlet velocities of water (0.001 m/s, 0.002 m/s, 0.003 m/s, 0.005 m/s and 0.01 m/s) and different material modules (pure PCM and hierarchical porous copper + PCM composite) were studied in the simulation. Abusive 2C-2C charge-discharge cycle was repeated for 3 times to test the performance of the BTMS under extreme battery operation scenario. The initial temperature was set to be 298 K. The Cu + paraffin composite was simulated as a single medium using the thermal conductivity, latent heat, and heat capacity obtained experimentally in the previous sections.

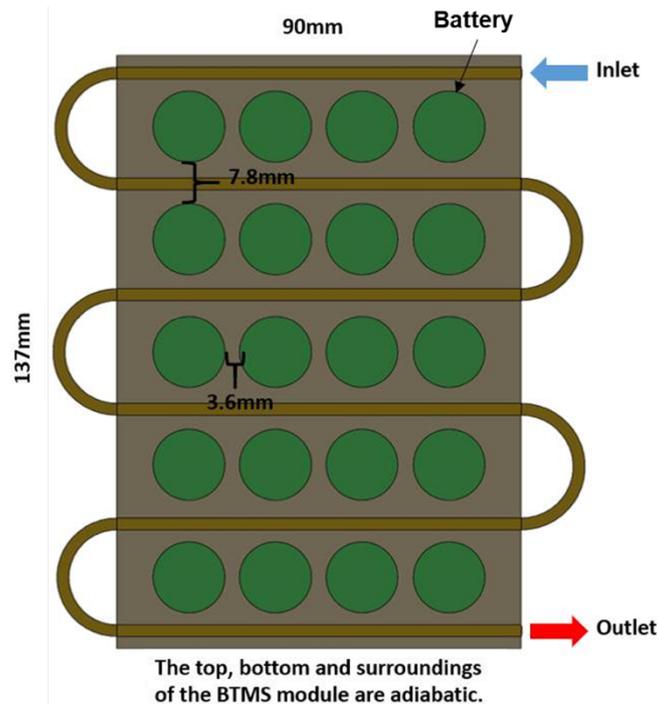

Figure 10. 2D model of the battery thermal management system encompassing 20 batteries surrounded by a PCM-filled array and a pipe in which a cooling liquid flows

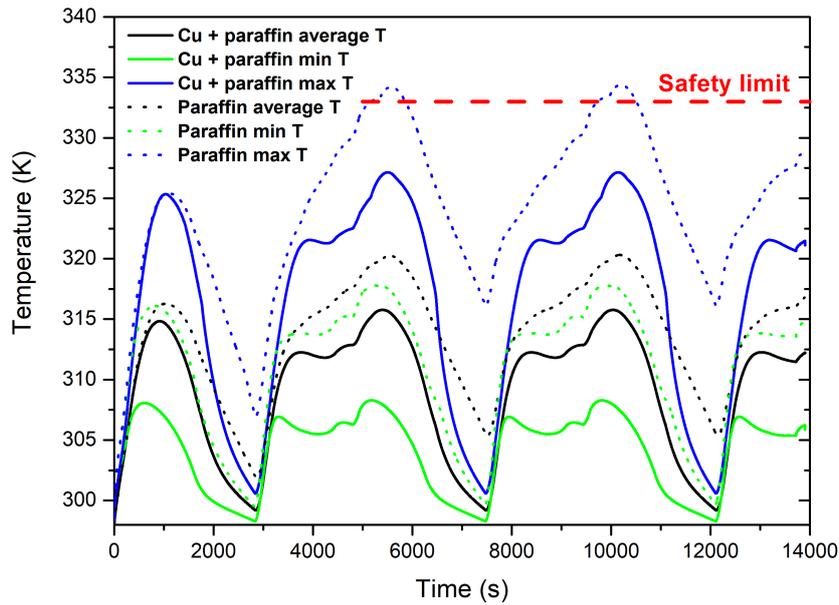

Figure 11. Maximum, minimum and average battery pack temperatures during three 2C-2C charge-discharge cycles, with inlet velocity of 0.002 m/s using Cu + paraffin or only paraffin for cooling

The results clearly indicate that Cu + paraffin outperforms pure paraffin in the first cycle, especially in the minimum temperature achieved by the batteries (Figure 11). However, and perhaps more importantly, with Cu + paraffin the battery pack remains below the temperature safety operation limit, 333 K, throughout the cycles, which was not achieved with pure paraffin and which could lead to thermal runaway. The difference between pure paraffin and Cu + paraffin increases after the first cycle and stabilizes after the second one, illustrating the great potential benefit that can be achieved adopting the hierarchical Cu-based shape-stabilized PCM proposed in this work.

Minimum and maximum battery temperatures with the composite were lower than with the pure PCM, due to significant thermal conductivity improvement. In particular, it can be seen that, when the composite was used, the maximum battery temperature is significantly lower compared to pure paraffin. At 7500 s a 15 K difference can be seen in Figure 11. The same trend holds for the average battery temperature and for the minimum battery temperature, which is 5 K and 10 K lower for the composite, respectively.

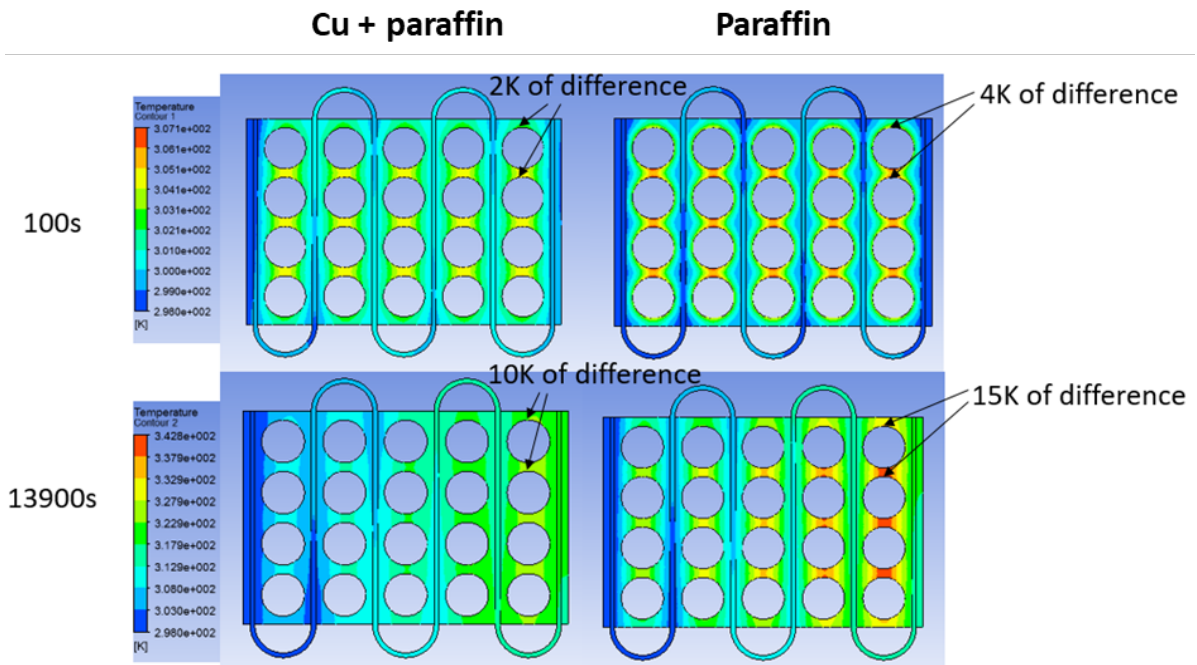

**Figure 12. Temperature map within the BTMS module with inlet velocity of 0.002 m/s and Cu + paraffin (left) or paraffin (right) for cooling at 100 s (top) and 13900 s (bottom) of battery operation**

Contour plots of the temperature further confirm that the Cu + paraffin composite performs better as compared to pure PCM when applied in BTMS. As shown in Figure 12, at the beginning (100 s) and close to the end (13900 s) of the battery charge/discharge process the temperature distribution was more homogenous with the composite compared to the case of pure PCM. In particular, at 100 s the maximum temperature difference for a single battery in the closest packing direction was 4 and 2 K for pure paraffin and composite, respectively. At 13900 s this difference grows to 15 and 10 K, confirming the higher ability of the composite system to guarantee a uniform temperature distribution as compared to pure paraffin. The maximum temperature difference for a single cell is depicted in Figure 13 as a function of the inlet velocity.

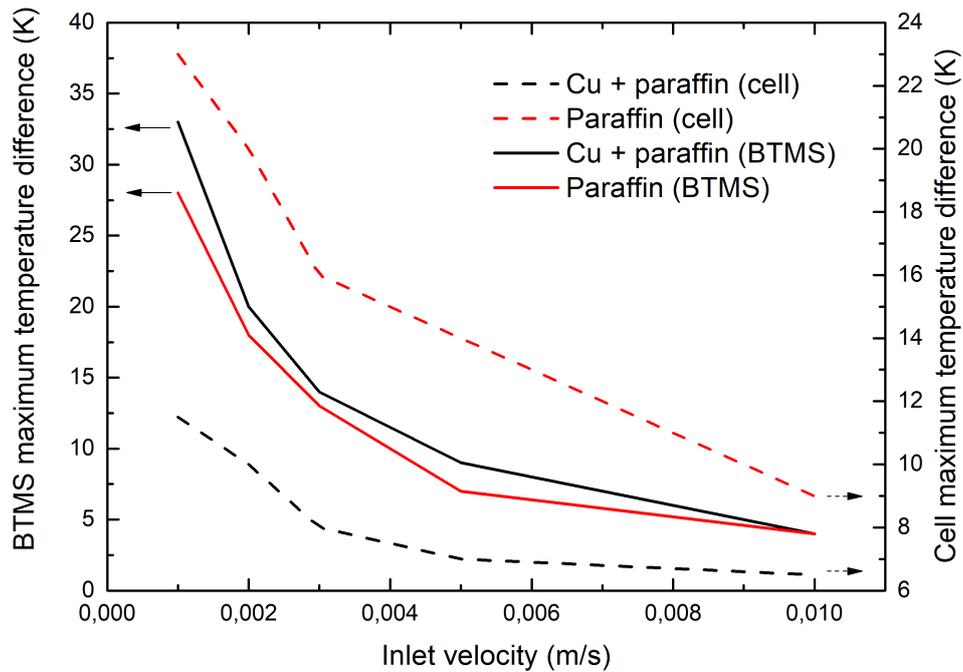

Figure 13. Variation of the maximum temperature difference within the battery pack (solid lines) and within a single cell (dashed lines) under different inlet velocities

Apart from the battery temperature itself, the temperature difference within singles cells is also crucial for a safe battery operation. In particular, temperature difference above 5 K can lead to the capacity mismatched for a single battery and even lead to a short circuit. As shown in Figure 11, Cu + paraffin composite maintains the maximum battery temperature under the safety limit when 0.002 m/s inlet velocity was used. However, there is a significant temperature difference within the cells (Figure 13), which is hazardous for the battery operation. This problem can be solved by higher inlet velocity of the cooling liquid. Under inlet velocity of 0.01 m/s the maximum battery temperature difference within cells was kept at 4 K (Figure 13), which is within the safe range. It is seen from Figure 13 that the maximum temperature difference with copper + PCM composite was higher than with pure PCM. This is due to the more efficient heat extraction from the battery pack, which accumulates in the cooling liquid. This results in a relative heat up of batteries close to the outlet. In other words: the overall temperature in the battery pack cooled by Cu + paraffin was lower, but the temperature gap in the system was slightly higher at low inlet velocities. This result is important in view of optimizing the operational conditions of the cooling system.

It should be noted that when liquid cooling is involved in a battery thermal management system, an increase of heat transfer fluid flow rate can only enhance heat transfer to a certain degree [95]. As can be observed from Figure 13 the inlet velocity of 0.01 m/s results in similar maximum temperature difference of battery module for the case of Cu + paraffin and for the case of pure paraffin. Therefore, it was not meaningful to explore higher velocities.

It should be noted that the improvements obtained in the thermal management of the battery pack come at the expense of a certain weight increase of the device. In particular, filling the empty space of the pack with paraffin increases the weight of the device by 36 %, while filling it with Cu + paraffin composite by 86 %. For more details on such weight calculation, please see Section 10 of the Supplementary information.

**Conclusions**

**In this work, for the first time, a new type of porous materials – hierarchical trimodal macro-nanoporous metals – was created, characterized, and proposed for shape stabilization, antileakage, and thermal conductivity enhancement of phase change materials (PCMs). In particular, trimodal hierarchical porous copper was synthesized by dealloying a Cu-Mg-Zn close-to-eutectic alloy, which was then impregnated with paraffin.**

**It was demonstrated experimentally as well as theoretically that such hierarchical macro-nanoporous materials have superior antileakage characteristics due to the presence of nanopores. It is worth remarking that, due to the presence of macropores, this goal was achieved without compromising the amount of PCM material that can be hosted in the porous media (around 90 vol%). This guarantees that our hierarchical trimodal macro-nanoporous metal system impregnated by paraffin PCM has a high energy density, very close to that of pure paraffin PCM. Additionally, it was shown that the use of the hierarchical macro-nanoporous metal system results in a threefold increase of the thermal conductivity of the shape-stabilized *vs* pure PCM. This enhances the rapidity of the response of the system to thermal *stimuli* and, thus, the overall efficiency of thermal storage or thermal control systems based on it, as shown in the thermal control of a battery pack application. By CFD modelling we have shown**

**that the proposed composite can be used for battery pack thermal management and, unlike pure paraffin, maintains the temperature of the battery within the safety limits.**

**The obtained results pave the way for hierarchical macro-nanoporous metal materials, synthesized by dealloying alloys of close-to-eutectic compositions, to be used for the development of high-energy density, leakage-free, and shape-stabilized PCMs with enhanced thermal conductivity. Such composites can be used for the effective thermal management of compact devices, like electronics or batteries and, more in general, for energy storage, which is a crucial building block for future distributed energy ecosystems.**

**Acknowledgement**

This work is supported by the Basque Government through its Ikermugikortasuna Mobility Grant for Research Collaboration number MV_2019_1_0020 as well as by UK EPSRC grants EP/S016627/1 (The Active Building Centre) and EP/P004709/1 (Energy-Use Minimisation via High Performance Heat-Power-Cooling Conversion and Integration). This project has received funding from the European Research Council (ERC) under the European Union's Horizon 2020 research and innovation programme (grant agreement No 803213). The help of Yagmur Polat with TG and DSC measurements as well as the help of María Jáuregui and Dr. Damien Saurel with XRD measurements are highly appreciated. The authors thank Dr. Alexander Lowe for his useful comments regarding the revised manuscript.